# Ultrafast core-to-core luminescence in BaF$_2$-LaF$_3$ single crystals

R. Shendrik[1], E. Radzhabov[1], A. Myasnikova[1], V. Pankratova[2], A. Šarakovskis[2], A. Nepomnyashchikh[1], A. Bogdanov[1], V. Gavrilenko[1] and V. Pankratov[2*]

[1] *Vinogradov Institute of Geochemistry, SB RAS, Irkutsk 664033, Russia*
[2] *Institute of Solid State Physics, University of Latvia, 8 Kengaraga, LV-1063 Riga, Latvia*

**ABSTRACT**. This study investigates the mechanisms underlying ultrafast cross-luminescence observed in BaF$_2$ crystals doped with LaF$_3$. We identified an ultrafast luminescent component with a decay time of approximately 150 ps, which emerges under excitation energies exceeding 24 eV as a novel *radiative* recombination process between electrons in the 5p core band of Ba$^{2+}$ and holes in the 5p core band of La$^{3+}$. Ab initio calculations support this hypothesis, showing that the energy levels of the core bands facilitate such transitions. The findings indicate that BaF$_2$-LaF$_3$ scintillators hold significant promise for applications in time-of-flight tomography.

## I. INTRODUCTION

Cross-luminescence (or core-valence luminescence) was discovered in BaF$_2$ crystals [1]. This luminescence has a short decay time constant of less than 1 ns and no measured rise time. The mechanism of cross-luminescence was proposed in [2–4]. Cross-luminescence occurs due to the excitation of core 5p-levels of Ba$^{2+}$ cations. During excitation, holes are formed in the core band, which can recombine with electrons in the valence band. The oscillator strength of these radiative transitions is high [5,6], resulting in a very short decay time constant.

Later, cross-luminescence was observed in other wide-bandgap compounds containing Cs, K, Rb, and Ba [3,7–9]. The gap between the top of the core band and the bottom of the valence band should be lower than the band gap of the compound for cross-luminescence to occur. Due to the short decay time constant, cross-luminescent materials are promising candidates for fast scintillators in high counting rate applications for muon detection in the mu2e project of Fermi lab [10] and other electromagnetic calorimeter modules [11].

Recently, it has been shown that BaF$_2$-based scintillators are promising for time-of-flight positron-emission tomography (TOF-PET) and computed tomography (TOF-CT) [12]. The critical challenge in TOF-PET is achieving a coincidence time resolution (CTR) of about 10 ps, which allows for a reduction in spatial resolution to 1 mm in real time. This is important for decreasing patient dose [13]. TOF-CT can increase the signal-to-noise ratio and reduce the dose rate by several orders of magnitude.

The cross-luminescence in BaF$_2$ crystals has one of the largest light output compared to other cross-luminescent materials [3]. Therefore, BaF$_2$ single crystals are promising for TOF tomography applications. A drawback of BaF$_2$-based scintillators is the presence of a long decay time component peaking at about 290 nm, which is attributed to self-trapped excitons (STE). However, it can be suppressed using special band-pass filters and metamaterials [14–17], although this method is ineffective due to light losses. An alternative suppression method has been proposed using a co-doping strategy. It has been found that the best reduction of STE luminescence occurs after doping BaF$_2$ with La$^{3+}$ [18–21], Cd$^{2+}$ [22] or Y$^{3+}$ [23] activators.

An unpolished sample of BaF$_2$-30 mol% LaF$_3$ without optical greasing demonstrates a CTR of 89 ± 3 ps. However, the CTR of polished sample with greasing should be reduced to 25 ps. This is one of the best CTR parameters for known TOF detectors. An ultrashort decay time of about 100 ps under 511 keV excitation was detected in La-doped samples. Its intensity increases with the concentration of La$^{3+}$ [24]. However, the mechanism of this luminescence is currently unknown.

In this article, we clarify the mechanism and present the time-resolved luminescence and excitation spectra of the ultrashort luminescence component.

## II. METHODS

### A. Spectroscopy

The time-resolved spectra were measured at the photoluminescence endstation (Finestlumi) [25,26] of FinEstBeAMS beamline [27,28] of MAX IV synchrotron facility at Lund (Sweden). The measurements were performed in the single bunch regime of the storage ring. The luminescence signals were registered using the Andor Shamrock (SR-303i) 0.3 m spectrometer coupled with the thermoelectrically cooled hybrid photodetector HPM-100-07C (Becker & Hickl).

---

* Corr. author: vladimirs.pankratovs@cfi.lu.lv

The intensity of each component in the decay is obtained through the deconvolution of the curves monitored at different wavelengths. The ideal decay time curve has a shape described by the following equation: $I_{ideal}(\lambda, t) = \sum_{i=0}^{N} A_i(\lambda)\exp(-t/\tau_i(\lambda))$. The real luminescence decay curve is $I_{real}(\lambda, t) = IRF(t) \otimes I_{ideal}(\lambda, t)$, where IRF means instrument response function. The intensity of each component at $\lambda$ is $I_i(\lambda) = A_i(\lambda)\tau_i(\lambda)$.

Raman spectra of crystals were obtained using a WITec alpha 300R confocal Raman spectroscopic system (WITec GmbH, Ulm, Germany) coupled with a 15 mW Nd:YAG laser ($\lambda$ = 532 nm). The spectra were recorded in the range from 150 to 1200 cm$^{-1}$ with diffraction grating (1800 g mm$^{-1}$) and spectral resolution of about 3 cm$^{-1}$.

Ab initio calculations for crystals of mixed barium halides were carried out within the framework of density functional theory using the VASP software package [29] on the "Akademik V.M. Matrosov" computing cluster [30]. For the calculations, a 2×2×2 supercell (96 atoms) was constructed in which one rare earth ion was substituted for one of the lattice cations. The atomic positions and crystal symmetry were obtained from the ICSD database [31]. The geometry optimization calculations were performed using the gradient approximation with the PBEsol exchange-correlation functional. Integration over the Brillouin zone was carried out using a G-centered grid of 8 k-points in the irreducible part of the Brillouin zone. The geometry optimization was conducted while preserving the shape and volume of the cell. Convergence was considered achieved if the difference in total energies between two iterations did not exceed 10$^{-6}$ eV. The energy cutoff for the plane-wave expansion was set at 500 eV.

Currently, periodic calculations using the density functional method are performed within the framework of the gradient approximation with the exchange-correlation potential PBE or PBEsol. It is known that using the PBE density functional for calculations in semiconductors and insulators leads to delocalized electronic states and, consequently, underestimated bandgap energies [32]. The most accurate methods of the GW0 type [33], which provide very good agreement with experimental results for halide crystals, are also gaining popularity, but these methods require substantial computational resources. Additionally, DFT calculations are conducted using hybrid functionals (PBE0, HSE), yielding good agreement with experimental data. In the previous works [34,35], it is shown that the hybrid functional PBE0 provides fairly good results comparable to experimental data and does not require large computing resources. In this work, we will use the PBEsol functional to calculate the equilibrium geometry of the lattice and PBE0 to calculate the density of states (DOS) of the crystal.

Calculations were performed for both a defect-free $BaF_2$ crystal and a crystal activated by $La^{3+}$ ions. Instead of cations, $La^{3+}$ ions were placed in amounts of 1 (corresponding to a concentration of ~3%), 2 (~6%), and 5 ions (~15%). The locations of the lanthanum ions were chosen randomly. Since the supercell as a whole must be electrically neutral, additional fluorine ions were placed in the nearest interstices to compensate for the excess charge of the trivalent ions.

**B. Crystal growth and characterization**

The $BaF_2$ single crystals doped with different concentrations of ultrapure and dried $LaF_3$ were grown by the Bridgman technique in a graphite crucible [20]. $CdF_2$ was added to the raw materials to avoid the incorporation of oxygen impurities into the crystal. The concentration of $La^{3+}$ in the $BaF_2$ single crystals was controlled using LA-ICP-MS.

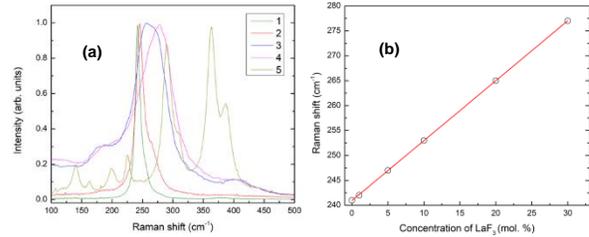

FIG. 1. Raman spectra of $BaF_2$ – $xLaF_3$ solid solutions, where x = 0 mol. % (curve 1), 5 mol% (curve 2), 20 mol% (curve 3) and 30 mol% (curve 4). For comparison, the Raman spectrum of $LaF_3$ single crystal is given (curve 5) (a) and the dependency with $LaF_3$ mole fraction of the Raman peak position of $BaF_2$ – $xLaF_3$ solid solutions (b).

$BaF_2 - xLaF_3$ (x=0-50 mol%) forms solid solution crystals with a fluorite structure and exhibits fluorine disordering when x > 10 mol% [36]. The Raman spectra of crystals with varying x values are shown in Fig. 1a. For comparison, the $LaF_3$ spectrum with a tysonite structure is also included. As the proportion of $LaF_3$ in the solid solution $BaF_2$-$xLaF_3$ increases, the Raman frequency of lattice vibrations rises. The full width at half maximum (FWHM) of the Raman bands of $BaF_2$-$xLaF_3$ crystals, where x > 10 mol%, is twice that of the undoped $BaF_2$ single crystal. This increase is attributed to fluorine disordering, as the fluorine anions require charge compensation for the $La^{3+}$ ions substituting for $Ba^{2+}$ cations. The band shift follows Vegard's law, suggesting that calcium and strontium are mixed homogeneously and randomly at the cation sites in the crystal with the fluorite structure (Fig. 1b).

## III. RESULTS AND DISCUSSION

The ultrafast component appears in the $BaF_2$-$LaF_3$ single crystals when excited with energies higher than 24 eV. A comparison of the decay time curves for nominally pure $BaF_2$ and $BaF_2$ with 30 mol% $LaF_3$ is presented in Fig. 2.

The observed short-lived luminescence, with a decay time constant of approximately 650 ps, corresponds to the regular component of cross-luminescence. However, it is evident that an ultrafast component appears in the luminescence of $BaF_2$-30 mol% $LaF_3$. Its intensity is about 1% of the total intensity, but the decay time constant is approximately 150 ps (see details in Table I).

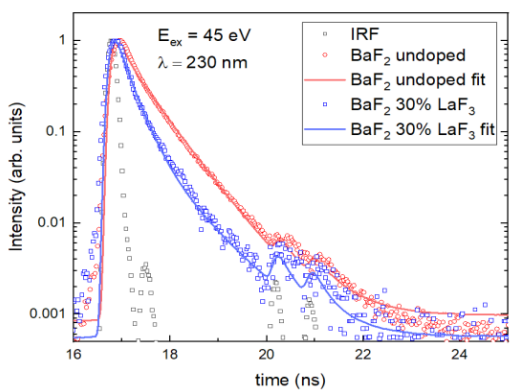

FIG. 2. Decay time curves of undoped $BaF_2$ and $BaF_2$-30 mol% $LaF_3$ under 45 eV excitation at 230 nm.

TABLE I. Decay time curve analysis of undoped $BaF_2$ and $BaF_2$-30 mol% $LaF_3$ under 45 eV excitation.

| | $\tau_1$, ns | $A_1$ (%) | $\tau_2$, ns | $A_2$ (%) | $\tau_3$, ns | $A_3$ (%) |
|---|---|---|---|---|---|---|
| $BaF_2$ | 230 nm | | | | | |
| | | | 0.62 | 89 | 10 | 11 |
| $BaF_2$ - 30 mol% $LaF_3$ | 230 nm | | | | | |
| | 0.15 | 0.7 | 0.77 | 99.3 | | |
| | 260 nm | | | | | |
| | 0.15 | 2 | 0.77 | 88 | 10 | 10 |

The time-resolved cross-luminescence emission spectra of the 150 ps and 650 ps components are given in Fig. 3. The 150 ps ultrafast luminescence time-resolved spectrum contains two bands peaking at approximately 245 nm and 270 nm. The fast 650 ps cross-luminescence band is shifted to a shorter wavelength region, peaking at about 235 nm. The STE luminescence band is located around 280 nm, with a decay time constant of about 10 ns due to concentration quenching of the STE luminescence in the $BaF_2$-30 mol% $LaF_3$ crystal.

The excitation spectra of the 150 ps and 650 ps components are presented in Fig. 4. The 650 ps component is excited at energies higher than the edge of 18 eV, which corresponds to the energy transition from the core 5p level of Ba to the 5d states that form the conduction band. The excitation edge for the 150 ps component is shifted to a higher energy region, located at approximately 23.5 eV.

FIG. 3. Time-resolved spectra of ultrafast 150 ps (blue curve) and fast 650 ps (red curve) components of cross-

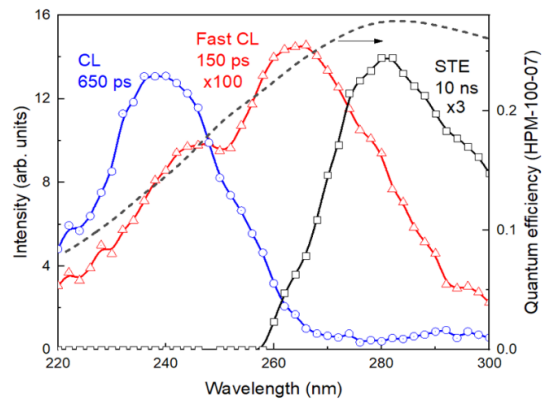

luminescence under 45 eV excitation in $BaF_2$-30 mol% $LaF_3$. The black curve shows STE-related luminescence. The dashed curve is the quantum efficiency of the HPM-100-07 photodetector.

The behavior of the excitation spectra for the 150 ps and 650 ps components differs. In the range of 50-1000 eV, the 650 ps component practically does not depend on energy. However, the 150 ps component demonstrates a monotonous increase. As shown in [24], the contribution of the ultrafast component increases to 20-30% of the total intensity under 511 keV excitation.

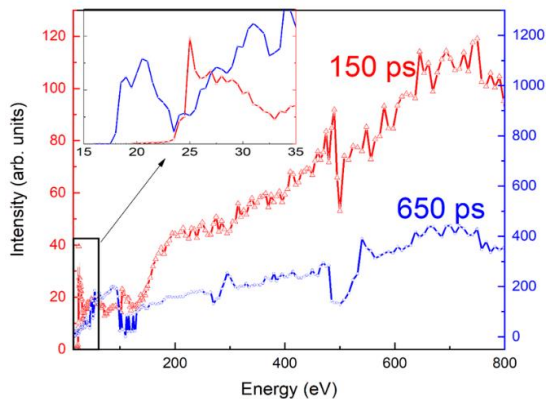

FIG. 4. Excitation spectra of the fast (650 ps) and ultrafast (150 ps) components of cross-luminescence monitored at 230 nm.

The calculation results are presented in Fig. 5. Initially, the parameters were calculated for a defect-free BaF$_2$ crystal. Ion displacements during geometry optimization turned out to be insignificant and amounted to no more than 1% of the lattice constant, according to experimental data of 6.2 Å [37]. The band gap of the crystal, estimated as the distance between the upper occupied and lower unoccupied states, was found to be 8.2 eV, which is about 1.2 times less than the experimental value. However, this method provides more accurate values than PBEsol, which underestimates the band gap by more than two times. The calculated density of states is shown in Fig. 5 (a). The valence band is predominantly formed by the 2p states of fluorine ions, while the upper core band is formed by the 5p states of barium ions. No additional levels are observed near the upper core band of a defect-free crystal, which corresponds to the classical concept of the band structure of barium fluoride.

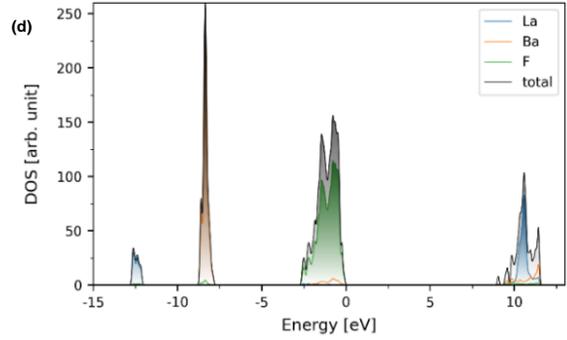

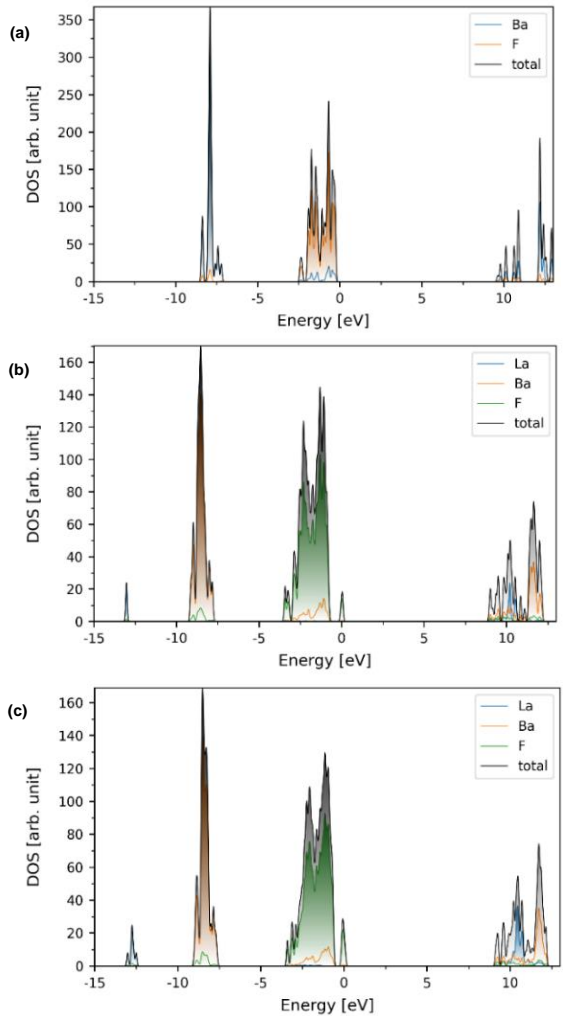

FIG 5. Density of States and Projected Density of States of BaF$_2$ (a) and BaF$_2$ – LaF$_3$ crystal with different concentrations: b) 3% LaF$_3$, c) 6% LaF$_3$, d) 15% LaF$_3$.

When a lanthanum atom is introduced into the lattice, local distortions occur, induced by both the impurity ion and the additional interstitial fluorine ion. In this scenario, two competing processes take place: the lattice anions shift closer to the trivalent lanthanum while simultaneously moving away from the interstitial fluorine ion. On average, the mixing of lattice anions and cations is no more than 8% of the distances found in a defect-free crystal. It is also noteworthy that the lanthanum and fluorine ions shift toward each other; the distance between them after relaxation was measured at 2.61 Å. In this case, the band structure of the crystal contains both the 5p levels of lanthanum ions, which lie below the core band, and the 2p levels of interstitial fluorine ions, which are in the forbidden band of the crystal (Fig. 5 (b, c, d)). As the La$^{3+}$ concentration increases, this level expands into the subband. The energy difference between the bottom of the 5p La$^{3+}$ band and the bottom of the conduction band is about 21 eV. The calculated distance between the bottom of the 5p Ba$^{2+}$ band and the bottom of the conduction band is about 16 eV.

Ab initio calculations show that the 5p La$^{3+}$ core states are located below the core band of Ba$^{2+}$. The calculated energy gap between the 5p levels of the Ba$^{2+}$ bottom of core band and the 2p F$^-$ band is about $E_{cv} = 5$ eV, and the distance between the bottom of core 5p Ba$^{2+}$ and core 5p La$^{3+}$ levels is about $E_{cc} = 4$ eV. Thus, the relationship $E_{cc} < E_{cv} < E_g$, where $E_g$ is calculated band gap (8.5 eV) is satisfied, indicating that core-to-core transitions between the core 5p levels of La$^{3+}$ and Ba$^{2+}$ could occur. The probability of these transitions increases with the widening of the La$^{3+}$ subband in crystals doped with high concentrations of La$^{3+}$. The position of the maximum of the $^5p_{3/2}$ level of La$^{3+}$ in the LaF$_3$ crystal is approximately 1.5 eV lower [38] than that of the $^5p_{3/2}$ level of Ba$^{2+}$ in the BaF$_2$ crystal [39], which is quite close to the calculations.

Assuming that the observed 150 ps luminescence is associated with core-core transitions between the 5p states of $Ba^{2+}$ and $La^{3+}$, a simplified model of this process is given in Fig. 6. Under excitation of the core 5p levels of $Ba^{2+}$ with energies higher than $E_{cv}^{exc} > 17$ eV, the regular 650 ps cross-luminescence occurs due to the recombination of electrons from the valence band and holes in the core $Ba^{2+}$ band ($h\nu_{cv}$). In $BaF_2$-$LaF_3$ crystals, the core 5p $La^{3+}$ band can also be excited; however, its excitation threshold should be higher.

The threshold energy for the excitation of the 150 ps luminescence is $E_{cc}^{exc} > 24$ eV, which is close to the excitation threshold of the core 5p-$La^{3+}$ band estimated from the calculations ($E_g^{c(La)} = 21$ eV) and higher than the $E_g^{c(Ba)} = 17$ eV excitation threshold of the regular 650 ps luminescence. Therefore, the 150 ps component can be attributed to the radiative recombination of electrons from the 5p $Ba^{2+}$ core band with holes in the 5p $La^{3+}$ core band ($h\nu_{cc}$). Core-to-core luminescence occurs. The distance between the 5p core bands of $Ba^{2+}$ and $La^{3+}$ ($E_{cc}$) is less than the distance between the 5p $Ba^{2+}$ core band and the 2p $F^-$ valence band ($E_{cv}$). Therefore, a redshift of the core-to-core luminescence can be expected. The width of $La^{3+}$ 5p core band ($\Delta E_c^{La} = 1.88$ eV) is less than the width of 5p $Ba^{2+}$ core band ($\Delta E_c^{Ba} = 2.56$ eV), following the calculation. Therefore, the hole relaxation process occurs faster in 5p $La^{3+}$ than in 5p $Ba^{2+}$ core bands [39,40]. This fact could explain the shorter decay time constant of core-to-core luminescence. The two peaks in the fast 150 ps core-to-core luminescence spectrum could arise due to spin-orbital coupling of 5p $Ba^{2+}$ upper core band on two core levels $^5p_{1/2}$ and $^5p_{3/2}$ [39] or due to transitions between the 5p-orbitals of $La^{3+}$ and the 2p orbitals of the nearest fluorine ions [6].

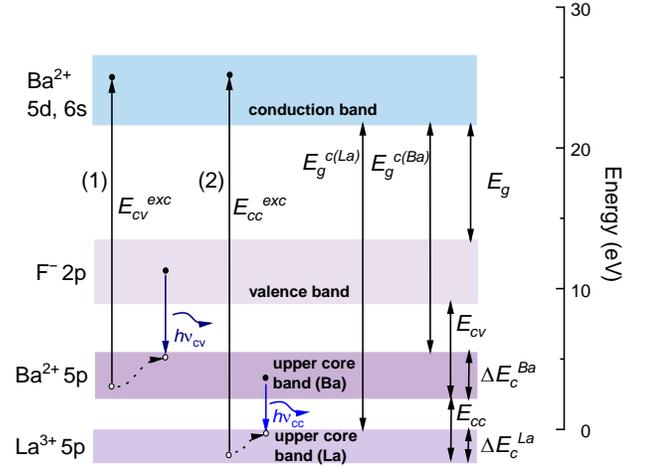

FIG. 6. The simplified mechanism of core-valence (650 ps) and core-to-core (150 ps) luminescence.

## IV. CONCLUSIONS

Ultrafast luminescence with a decay time constant of 150 ps in $BaF_2$-$LaF_3$ crystals has been observed for the first time. It exhibits a 1 eV redshift compared to the regular 650 ps cross-luminescence. The ultrafast 150 ps luminescence has an excitation threshold at 24 eV. Based on the ab initio calculations, we propose that this luminescence could be related to the core-to-core recombination of an electron in the 5p $Ba^{2+}$ band and a hole in the 5p $La^{3+}$ band.

## ACKNOWLEDGMENTS

The authors gratefully acknowledge Dr. Kirill Chernenko from MAX IV Laboratory for his assistance in synchrotron experiments. V.P. is thankful for the financial support from Latvian Research Council, Grant number LZP-2022/1–0611 "Cross-luminescence engineering for picosecond time-of-flight gamma and X-ray imaging for medical applications". A.Š. acknowledges state program VPP-EM-FOTONIKA-2022/1-0001. The research was also supported by the project 0284-2021-0004 (Materials and Technologies for the Development of Radiation Detectors, Luminophores, and Optical Glasses). Raman spectra were measured at the Center for Geodynamics and Geochronology of the Institute of Earth's Crust SB RAS.